\documentclass{article}
\usepackage{arxiv}

\usepackage[utf8]{inputenc} % allow utf-8 input
\usepackage[T1]{fontenc}    % use 8-bit T1 fonts
\usepackage{hyperref}       % hyperlinks
\usepackage{url}            % simple URL typesetting
\usepackage{booktabs}       % professional-quality tables
\usepackage{amsfonts}       % blackboard math symbols
\usepackage{nicefrac}       % compact symbols for 1/2, etc.
\usepackage{microtype}      % microtypography
\usepackage{lipsum}
\usepackage[version=3]{mhchem} % Formula subscripts using \ce{}
\usepackage{mathptmx}
\usepackage{multirow}
\usepackage{hyperref}
\usepackage{braket}
\usepackage{rotating}
\usepackage{color}
\usepackage{graphicx}
\usepackage{amsmath}
\usepackage[square,numbers,compress]{natbib}
\usepackage{notes2bib}
\usepackage{comment}
\usepackage[symbol]{footmisc}

\newcommand{\fig}{Fig.}
\newcommand{\Fig}{Fig.}
\newcommand{\figref}[1]{\fig~\ref{#1}}
\newcommand{\Figref}[1]{\Fig~\ref{#1}}

\newcommand{\tabref}[1]{table~\ref{#1}}

\renewcommand{\eqref}[1]{Eq.~(\ref{#1})}

\newcommand{\secref}[1]{Section~\ref{#1}}

\title{Transfer learning for chemically accurate interatomic neural network potentials$^\dag$}

\author{      Viktor Zaverkin$^1$\\
              University of Stuttgart\\
              Faculty of Chemistry\\
              Institute for Theoretical Chemistry\\
              \And
              David Holzmüller$^2$\\
              University of Stuttgart\\
              Faculty of Mathematics and Physics\\
              Institute for Stochastics and Applications\\
              \And
              Luca Bonfirraro\\
              University of Stuttgart\\
              Faculty of Chemistry\\
              Institute for Theoretical Chemistry\\
              \And
              Johannes Kästner\\
              University of Stuttgart\\
              Faculty of Chemistry\\
              Institute for Theoretical Chemistry\\
              }

\begin{document}

\footnotetext{$^{1}$~E-mail: \texttt{zaverkin@theochem.uni-stuttgart.de}}
\footnotetext{$^{2}$~E-mail: \texttt{david.holzmueller@mathematik.uni-stuttgart.de}}
\footnotetext{\dag~Electronic Supplementary Information (ESI) available. See DOI: 00.0000/00000000.}

\maketitle

\begin{abstract}
Developing machine learning-based interatomic potentials from \textit{ab-initio} electronic structure methods remains a challenging task for computational chemistry and materials science. This work studies the capability of transfer learning, in particular discriminative fine-tuning, for efficiently generating chemically accurate interatomic neural network potentials on organic molecules from the MD17 and ANI data sets. We show that pre-training the network parameters on data obtained from density functional calculations considerably improves the sample efficiency of models trained on more accurate \textit{ab-initio} data. Additionally, we show that fine-tuning with energy labels alone can suffice to obtain accurate atomic forces and run large-scale atomistic simulations, provided a well-designed fine-tuning data set. We also investigate possible limitations of transfer learning, especially regarding the design and size of the pre-training and fine-tuning data sets. Finally, we provide GM-NN potentials pre-trained and fine-tuned on the ANI-1x and ANI-1ccx data sets, which can easily be fine-tuned on and applied to organic molecules.
\end{abstract}

% keywords can be removed

\section{Introduction}

The impact of machine learning (ML) on chemical and materials science is tremendous as it extends the computationally affordable time and length scales when modeling and predicting physical and chemical phenomena~\cite{Dral2020, Mueller2020, Unke2021, Manzhos2021, Deringer2021}. Particularly, ML allows for constructing potential energy surfaces (PES) with a computational efficiency comparable to classical force fields and an accuracy on par with \textit{first-principles} methods. However, some applications in computational chemistry and materials science require electronic structure methods with accuracy far beyond the conventionally used density functional theory (DFT). For example, \textit{ab-initio} methods, such as coupled-cluster theory~\cite{Purvis1982, Crawford2000, Bartlett2007}, systematically approach the exact solution of the Schr{\"o}dinger equation and provide chemically accurate total energies and atomic forces. At the same time, the data set sizes accessible at the respective level of theory are often limited due to the high computational cost, while calculating atomic forces can be infeasible.

Given a sparse data set at a chemically accurate level of theory, a supervised ML methods' data efficiency is central for developing reliable interatomic potentials. Different approaches can be used for this purpose. For example, our earlier works developed ensemble-free active learning approaches for interatomic neural network (NN) potentials based on the last layer and sketched gradient features~\cite{Zaverkin2021a, Zaverkin2022d, Holzmueller2022}. These approaches provided a learned similarity measure between data points by considering the gradient kernel of a trained NN, which corresponds to the finite-width neural tangent kernel~\cite{Jacot2018}. Additionally, to avoid selecting similar structures, Refs.~\citenum{Zaverkin2022d, Holzmueller2022} presented several methods that enforce the diversity and representativeness of the selected batch.

Several alternative approaches to Refs.~\citenum{Zaverkin2021a, Zaverkin2022d, Holzmueller2022} that employ ensembles or are ensemble-free can be found in the literature~\cite{Gastegger2017, Janet2017, Podryabinkin2017, Smith2018,  Nandy2018, Gubaev2018, Janet2019, Schran2020, Zhu2022}. Alternatively, one could augment the energy labels by using atomic forces~\cite{Cooper2020}, but this approach may be limited by the computational expense of computing the respective labels. Lastly, one can leverage information from larger data sets computed at a cheaper level of theory, such as DFT, through transfer learning.

Transfer learning is actively used for natural language processing~\cite{Howard2018, Devlin2019, Brown2020, Wei2022} and computer vision~\cite{Chen2020, He2020} tasks and achieves remarkable successes in these domains. Another field with increased interest in transfer learning is the drug discovery domain~\cite{Hu2020, Wu2021, Xie2022, Sun2022}. In this work, we are interested in investigating the application of transfer learning approaches to modeling interatomic interactions by artificial NNs. In the investigated transfer learning setting, the parameters of a model are first trained on the source task, for example, using DFT energy and atomic force labels. Then, the respective parameters are fine-tuned using the target data set, e.g., coupled-cluster labels. One of the main advantages of transfer learning is the improved data efficiency on the target task due to pre-training features on the source task.

An alternative approach to transfer learning, frequently used for training highly accurate interatomic ML potentials, is $\Delta$-learning~\cite{Ramakrishnan2015}. In this approach, a difference from a computationally cheap \textit{first-principles} method and an accurate \textit{ab-initio} method is learned by the respective ML approach. This approach requires running two models simultaneously during the inference step. Thus, $\Delta$-learning is somewhat less computationally efficient than transfer learning but provides the advantage of having a method that may preserve the model from escaping physically meaningful regions. An alternative approach to $\Delta$-learning would train an interatomic ML potential on the respective computationally cheap \textit{first-principles} method and train another ML potential on the difference from the former to an accurate \textit{ab-initio} method~\cite{Batra2019, Zaspel2019, Dral2020b}. Such approaches improve on the disadvantage of $\Delta$-learning on using computationally inefficient \textit{first-principles} method during simulations. For a more detailed discussion about transfer learning and various alternatives, we refer to Ref.~\citenum{Dral2023}. We only consider transfer learning approaches applied to interatomic NN potentials in this work and leave $\Delta$-learning for future work.

To the best of our knowledge, the application of transfer learning approaches to modeling interatomic NN potentials is mainly covered by Ref.~\citenum{Smith2019} and recently published Refs.~\citenum{Kaeser2020, Zheng2021, Kaeser2022, Gardner2022, Zhang2022, Chen2022, Gao2022}. While in Refs.~\citenum{Smith2019, Kaeser2020, Zheng2021, Kaeser2022, Gardner2022, Zhang2022, Chen2022} some hidden layers have been fine-tuned, the approach proposed in Ref.~\citenum{Gao2022} employs linear probing, i.e., it re-uses all parameters but not the last layer which is re-initialized. For most literature approaches~\cite{Smith2019, Zheng2021, Gardner2022, Zhang2022, Gao2022}, the resulting performance of the employed model has been evaluated only with respect to standard error measures as mean absolute (MA) or root-mean-squared (RMS) errors in total energies and atomic forces. In Ref.~\citenum{Chen2022}, the developed models were applied to simulate bulk liquid water at various \textit{ab-initio} levels of theory. Refs.~\citenum{Kaeser2020} and \citenum{Kaeser2022} employed transfer learning to investigate vibrational degrees of freedom of the \ce{H2CO} molecule and to determine chemically accurate tunneling splittings, respectively.

In this work, we propose an alternative approach to Refs.~\citenum{Smith2019, Gardner2022, Zhang2022, Gao2022, Chen2022} which utilizes discriminative fine-tuning~\cite{Howard2018}. Discriminative fine-tuning has been used previously in the natural language processing domain and allows adjustment of the fully-connected layers of an NN to a different extent, as they may require different amounts of adaptation. We employ the Gaussian moment neural network (GM-NN) approach~\cite{Zaverkin2020, Zaverkin2021b}, developed by some of us, to model interatomic interactions. Thus, we design the respective transfer learning approach to fit a framework with trainable representations and trainable atomic scale and shift parameters.

We thoroughly investigate the improvement in the model's data efficiency achieved by transfer learning. Particularly, we assess the model's performance in predicted atomic forces, as they are essential for most atomistic simulations, while fine-tuning the respective model using energy or energy and atomic force labels. Note that we do not expect models fine-tuned on energy labels only to outperform those fine-tuned on energies and atomic forces. However, the former may provide an improved accuracy compared to models trained from scratch on energies and provide the means of generating reliable interatomic potentials for systems for which atomic forces are inaccessible at the desired level of theory. For this purpose, we employ two benchmark data sets, MD17~\cite{Chmiela2017, Schuett2017_2, Chmiela2018, Sauceda2019, Christensen2020b} and ANI~\cite{Smith2018, Smith2019, Smith2020}. We find that selected data set sizes for pre-training and fine-tuning can influence the final accuracy of developed potentials.

Moreover, we run molecular dynamics simulations on different molecular and bulk systems to more rigorously assess the quality of fine-tuned interatomic potentials. The investigated systems are the aspirin molecule and deca-alanine (Ala$_{10}$) in the gas phase and water. In addition, we highlight advantages and shortcomings of transfer-learned interatomic NN potentials, drawing particular attention to the design of fine-tuning data sets. In summary, we extend the observations of Refs.~\citenum{Smith2019, Kaeser2020, Zheng2021, Kaeser2022, Gardner2022, Zhang2022, Gao2022, Chen2022} regarding the improvement in sample-efficiency by also studying smaller data sets, differences between force-and-energy and energy-only fine-tuning, the effect of the pre-training set size, and the behavior in molecular dynamics simulations.

The software employed in this work is implemented within the Tensorflow framework~\cite{Abadi2015} and is available free-of-charge at \href{https://gitlab.com/zaverkin\_v/gmnn}{gitlab.com/zaverkin\_v/gmnn}, including the proposed transfer learning approach. The ANI interatomic potentials, obtained by pre-training and fine-tuning, will be published at \href{https://doi.org/10.18419/darus-3299}{doi.org/10.18419/darus-3299}.

The presented work is structured as follows: First, \secref{sec:methods} introduces the architecture of GM-NN-based potentials~\cite{Zaverkin2020, Zaverkin2021b} and describes the proposed transfer learning approach. \secref{sec:results} demonstrates the performance of the proposed transfer learning approach on selected benchmark systems. Finally, \secref{sec:discussion_conclusion} discusses and concludes this work's main findings, including limitations of transfer learning approaches applied to interatomic NN potentials.

\section{Methods \label{sec:methods}}

\begin{figure*}[t]
    \centering
    \includegraphics[width=16cm]{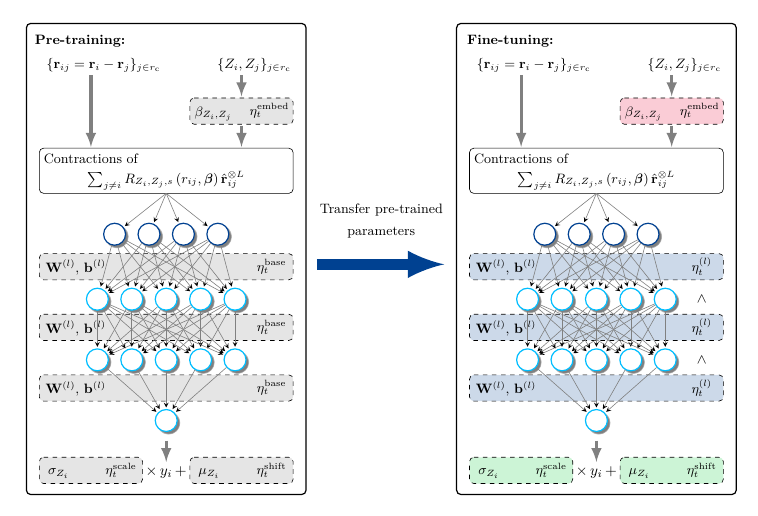}
    \caption{Schematic overview of the proposed transfer learning approach. Grey boxes denote parameters learned during the pre-training step with the respective learning rates $\eta_t$, i.e., embeddings $\boldsymbol{\beta}$ ($\eta_t^\mathrm{embed}$), weights $\mathbf{W}^{(l)}$ and biases $\mathbf{b}^{(l)}$ of the fully connected layers ($\eta_t^\mathrm{base}$), as well as atomic scale and shift parameters $\boldsymbol{\sigma}$ ($\eta_t^\mathrm{scale}$) and $\boldsymbol{\mu}$ ($\eta_t^\mathrm{shift}$). Red boxes denote parameters fixed during the fine-tuning step ($\boldsymbol{\beta}$), while blue boxes indicate that the parameters ($\mathbf{W}^{(l)}$ and $\mathbf{b}^{(l)}$) are fine-tuned during the fine-tuning step. We employ discriminative fine-tuning~\cite{Howard2018}, i.e., we fine-tune different layers to different extents by selecting different learning rates ($\eta^{(l+1)}>\eta^{(l)}$). Trainable atomic scale and shift parameters ($\boldsymbol{\sigma}$ and $\boldsymbol{\mu}$) are re-initialized at the beginning of and re-trained during the fine-tuning step, which is indicated by the green color.}
    \label{fig:scheme}
\end{figure*}

The following section presents the proposed transfer learning approach as illustrated in \figref{fig:scheme}. Throughout this work, we employ a particular architecture of interatomic NN potentials, i.e., the Gaussian moment neural network (GM-NN) approach~\cite{Zaverkin2020, Zaverkin2021b}. Thus, we begin this section with a brief review of the respective method in \secref{sec:gmnn}. The transfer learning approach is described in \secref{sec:tl}.

Additionally, to simplify the notation in the following, we define an atomic structure by $S = \{\mathbf{r}_i, Z_i\}_{i=1}^{N_\mathrm{at}}$ with $\mathbf{r}_i \in \mathbb{R}^3$ and $Z_i \in \mathbb{N}$ being the Cartesian coordinates and the atomic number of atom $i$, respectively. We consider learning the parameterized mapping of an atomistic structure to scalar electronic energy, i.e., $f_{\boldsymbol{\theta}}: S \mapsto E \in \mathbb{R}$. The mapping is learned from data $\mathcal{D} = \left(\mathcal{X}_\mathrm{train}, \mathcal{Y}_\mathrm{train}\right)$ with $\mathcal{X}_\mathrm{train} = \{S_k\}_{k=1}^{N_\mathrm{train}}$ and $\mathcal{Y}_\mathrm{train} = \big\{E_k^\mathrm{ref}, \{\mathbf{F}_{i,k}^\mathrm{ref}\}_{i=1}^{N_\mathrm{at}}\big\}_{k=1}^{N_\mathrm{train}}$. In general, different \textit{first-principles} or \textit{ab-initio} electronic structure methods can be used to compute the reference energy $E_k^\mathrm{ref}$ and atomic forces $\{\mathbf{F}_{i,k}^\mathrm{ref}\}_{i=1}^{N_\mathrm{at}}$.

\subsection{Gaussian moment neural network \label{sec:gmnn}}

To achieve linear scaling of the interatomic NN potentials' computational cost with  the number of atoms $N_\mathrm{at}$, we 
% To allow for developing interatomic NN potentials, whose computational cost scales linearly with the number of atoms $N_\mathrm{at}$, we
assume the locality of interatomic interactions, defined by a finite cutoff radius $r_\mathrm{c}$. Employing this approximation, the total energy of an atomistic system $S$ can be split into its atomic contributions~\cite{Behler2007}
\begin{equation}
	\label{eq:energy_model}
    E \left(S, \boldsymbol{\theta}\right) \approx \sum_{i=1}^{N_\mathrm{at}} E_{i}\left( \mathbf{G}_i, \boldsymbol{\theta} \right).
\end{equation}
Here, the neighborhood of an atom $i$ is encoded by a local atomic representation $\mathbf{G}_i$, referred to as Gaussian moment (GM)~\cite{Zaverkin2020}, which includes all necessary invariances and ensures efficient training of an atomistic NN. The GM representation is constructed by defining the pair distance vectors $\mathbf{r}_{ij} = \mathbf{r}_i - \mathbf{r}_j\,\forall\,j \in r_\mathrm{c}$ and splitting them into their radial and angular components, i.e., $r_{ij} = \lVert \mathbf{r}_{ij} \rVert_2$ and $\hat{\mathbf{r}}_{ij} = \mathbf{r}_{ij} / r_{ij}$, respectively. A representation that is equivariant to rotations can then be obtained as~\cite{Zaverkin2020, Zaverkin2021b}
\begin{equation}
    \boldsymbol{\Psi}_{i, L, s} = \sum_{j \neq i} R_{Z_i, Z_j, s}\left(r_{ij}, \boldsymbol{\beta}\right)  \hat{\mathbf{r}}_{ij}^{\otimes L},
\end{equation}
where $\hat{\mathbf{r}}_{ij}^{\otimes L} = \hat{\mathbf{r}}_{ij} \otimes \cdots \otimes \hat{\mathbf{r}}_{ij}$ is the $L$-fold outer product of the angular components and $R_{Z_i, Z_j, s}\left(r_{ij}, \boldsymbol{\beta}\right)$ are nonlinear radial functions with trainable parameters $\boldsymbol{\beta}$. The latter introduces species dependence in the employed representation. As radial functions, we employ a weighted sum of Gaussian functions~\cite{Zaverkin2021b} and re-scale them by the cosine cutoff function~\cite{Behler2007}, to ensure smooth dependence on the number of neighboring atoms. To obtain features invariant to rotations, we compute full tensor contractions of $\boldsymbol{\Psi}_{i, L, s}$ and employ unique generating graphs to eliminate possible linear dependencies~\cite{Zaverkin2020, Zaverkin2021b}.

To map the invariant features $\mathbf{G}_i$ to the scalar atomic energy $E_i$, we employ a fully-connected feed-forward NN consisting of two hidden layers~\cite{Zaverkin2021b}; see \figref{fig:scheme}. Our network consists of $360$ input neurons, $512$ hidden neurons in both hidden layers, and a single output neuron. All weights $\mathbf{W}^{(l)}$ and biases $\mathbf{b}^{(l)}$ are shared across all species as the corresponding alchemical information is encoded in $\mathbf{G}_i$. The weights are initialized by picking the respective entries from a normal distribution with zero mean and unit variance. The bias vectors are initialized to zero. Moreover, we employ the neural tangent parameterization to improve training efficiency and accuracy~\cite{Jacot2018}. The Swish/SiLU activation function is used as a non-linearity~\cite{Elfwing2018, Ramachandran2017}. Additionally, we employ trainable, species-dependent shift and scale parameters $\mu_{Z_i}$ and $\sigma_{Z_i}$
\begin{equation}
E_{i}\left(\mathbf{G}_i, \boldsymbol{\theta} \right) = c \cdot (\sigma_{Z_i} y_i + \mu_{Z_i}),
\end{equation}
to aid the training process. Here, $y_i$ is the direct output of the interatomic NN. The constant $c$ is defined as the root-mean-square (RMS) error per atom of the mean atomic energy estimated from the reference energy labels, $\mu_{Z_i}$ are initialized by solving a linear regression problem~\cite{Zaverkin2021b}, and $\sigma_{Z_i}$ are initialized to 1.

The parameters $\boldsymbol{\theta}$ of the NN, i.e., $\mathbf{W}$ and $\mathbf{b}$, as well as $\boldsymbol{\beta}$ of the local representation and the parameters $\sigma_{Z}$ and $\mu_Z$ that scale and shift the output of the NN are trained, i.e., optimized by minimizing the mean squared loss on training data
\begin{equation}
    \label{eq:loss}
    \begin{split}
        \mathcal{L}\left(\left. \boldsymbol{\theta} \right| \mathcal{D}\right) = \sum_{k=1}^{N_\mathrm{Train}} & \Bigg[\lambda_E \lVert E_k^\mathrm{ref} - E(S_k, \boldsymbol{\theta})\rVert_2^2 +  \\ & \quad  \lambda_F \sum_{i=1}^{N_\mathrm{at}^{(k)}} \lVert \mathbf{F}_{i,k}^\mathrm{ref} - \mathbf{F}_i\left(S_k, \boldsymbol{\theta}\right)\rVert_2^2\Bigg],
    \end{split}
\end{equation}
where $\lambda_E$~au and $\lambda_F$ have to be chosen to balance the energy and force loss contributions. For data sets consisting of equally sized structures, we employ $\lambda_E = 1$~au and $\lambda_F= 4$~au~\AA$^2$. If the employed data set contains configurations of different sizes, i.e., ANI-1x and ANI-1ccx in this work~\cite{Smith2018, Smith2019, Smith2020}, $\lambda_E = 1/N_\mathrm{at}$~au and $\lambda_F= 0.01$~au~\AA$^2$ are used. The atomic force of atom $i$ is defined as the negative gradient of the total energy with respect to the atomic position $\mathbf{r}_i$, i.e., $\mathbf{F}_i\left(S_k, \boldsymbol{\theta}\right) = -\nabla_{\mathbf{r}_i} E\left(S_k, \boldsymbol{\theta}\right)$.

The combined loss function in \eqref{eq:loss} is minimized by employing the Adam optimizer~\cite{Adam2015}. The respective parameters of the optimizer are $\beta_1=0.9$, $\beta_2=0.999$, and $\epsilon=10^{-7}$. We employ mini-batches of 32 and 2048 molecules for MD17~\cite{Chmiela2017, Schuett2017_2, Chmiela2018} and ANI data sets~\cite{Smith2020}, respectively. The presented work uses different layer-wise learning rates for the pre-training and fine-tuning steps in \figref{fig:scheme}; for more details, see \secref{sec:tl}. In a general setting, we employ learning rates of $\eta_0^{\mathrm{base}}=0.03$ for the parameters of the fully connected layers, $\eta_0^{\mathrm{embed}}=0.02$ for the trainable representation, as well as $\eta_0^{\mathrm{shift}}=0.05$ and $\eta_0^{\mathrm{scale}}=0.001$ for the shift and scale parameters of atomic energies, respectively. All learning rates are decayed linearly to zero by multiplying them with $(1 - t)$, i.e., $\eta_t = \eta_0 (1 - t)$, where $t =
\mathrm{step}/\mathrm{max\_step}$. The training is performed for 1000 training epochs for the MD17 data and 2000 steps for ANI. To prevent overfitting during training, we employed the early stopping technique~\cite{Prechelt2012}.

All models are trained within the Tensorflow framework~\cite{Abadi2015} on a central processing unit (CPU) node equipped with two Intel Xeon E6252 Gold (Cascade Lake) CPUs.

\subsection{Transfer learning interatomic potentials \label{sec:tl}}

In this section, we describe the employed transfer learning pipeline, which includes the pre-training and fine-tuning steps as depicted in \figref{fig:scheme}. Here we are most interested in the general transfer learning setting~\cite{Howard2018}, applied to interatomic potentials. We consider a larger source task $\mathcal{D}_\mathrm{S}$, which contains a set of atomic configurations and respective labels, e.g., energies and forces, and a smaller target task $\mathcal{D}_\mathrm{T}$, i.e., $N_\mathrm{S} = \mathrm{len} \mathcal{D}_\mathrm{S} > N_\mathrm{T} = \mathrm{len} \mathcal{D}_\mathrm{T}$. Transfer learning aims to improve the performance on $\mathcal{D}_\mathrm{T}$ by employing the structural information learned from $\mathcal{D}_\mathrm{S}$. Note that we allow the configurations in $\mathcal{D}_\mathrm{S}$ and $\mathcal{D}_\mathrm{T}$ to overlap, i.e., $\mathcal{X}_\mathrm{T} \subseteq \mathcal{X}_\mathrm{S}$. Importantly, we compute labels in $\mathcal{D}_\mathrm{S}$ and $\mathcal{D}_\mathrm{T}$ using different electronic structure techniques, e.g., DFT and CCSD(T)/CBS~\cite{Hobza2002, Feller2006}. Therefore, both tasks are generally closely aligned and thus may allow for the effective transfer of learned structural and alchemical information~\cite{Sun2022}.

The pre-training step in \figref{fig:scheme} uses the default setup of GM-NN models described in \secref{sec:gmnn}, including the layer-wise learning rates. In the context of transfer learning, the main benefit of pre-training is computing a better initialization of the model's trainable parameters than randomly initializing them. Having seen a larger number of atomic configurations, the interatomic NN potential model may capture better vibrational and compositional degrees of freedom, which are not present in the target tasks with a smaller amount of data. Thus, pre-training may lead to better convergence and generalization for tasks with fewer labeled samples. Note that pre-training, neglecting the acquisition of labels from \textit{ab-initio} calculations, is computationally the most expensive step but has to be performed only once.

\begin{figure*}[t!]
    \centering
    \includegraphics[width=16cm]{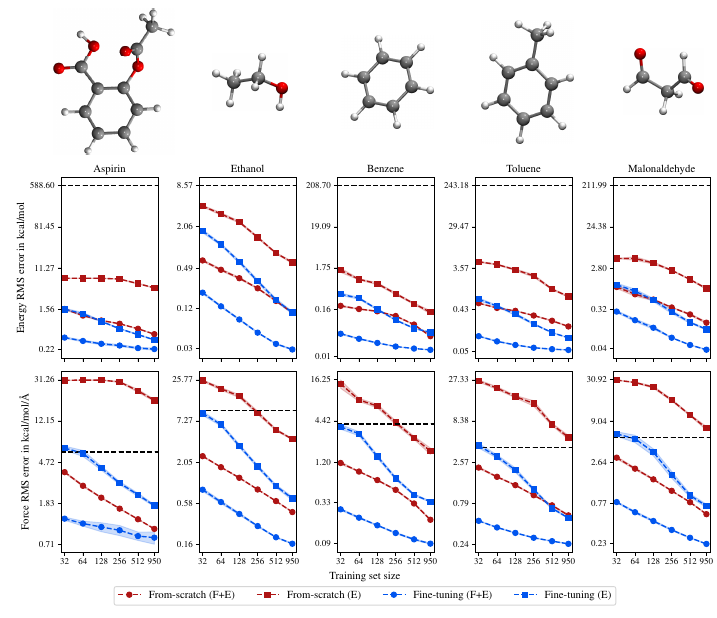}
    \caption{Learning curves for five molecules from the MD17 data set with the respective molecular geometries shown as an inset. The root-mean-square (RMS) errors of total energies and atomic forces are plotted against the training set size. Shaded areas denote the standard error of the mean evaluated over five independent runs. Models trained directly on the coupled-cluster data are referred to as ``from-scratch''. Models obtained by employing the fine-tuning approach to transfer from density functional-based pre-trained models are denoted by ``fine-tuning''. Simultaneous training on energy and atomic force information (E+F) is compared to energy-based training (E). The horizontal black line denotes the final accuracy of models trained on DFT and evaluated on coupled-cluster energy and atomic force labels.}
    \label{fig:md17_learning_curves_with_molecules}
\end{figure*}

The fine-tuning step in \figref{fig:scheme} is required as data used for pre-training uses different labels and thus may come from a different distribution. To the best of our knowledge, the application of transfer learning approaches to modeling interatomic NN potentials is mainly covered by Ref.~\citenum{Smith2019} and recently published Refs.~\citenum{Kaeser2020, Zheng2021, Kaeser2022, Gardner2022, Zhang2022, Chen2022, Gao2022}, while their application in drug discovery is somewhat broader~\cite{Hu2020, Wu2021, Xie2022, Sun2022}. In this work, we propose an alternative approach to Refs.~\citenum{Smith2019, Kaeser2020, Zheng2021, Kaeser2022, Gardner2022, Zhang2022, Gao2022, Chen2022} and investigate the performance of interatomic NN models, which have been trained only on energy labels during the fine-tuning step, on atomic forces.

We empirically found that the trainable parameters $\boldsymbol{\beta}$ of the descriptor should be fixed during the fine-tuning step, i.e., $\eta_0^{\mathrm{embed}} = 0.0$; see \figref{fig:scheme}. This might be because the pre-training already produces a good representation and fine-tuning it can easily lead to overfitting. The trainable scale and shift parameters, i.e. $\sigma_{Z_i}$ and $\mu_{Z_i}$, have to be re-initialized to account for possible differences in energy labels as, e.g., differences in cohesive and total energies. We use the default learning rates for $\sigma_{Z_i}$ and $\mu_{Z_i}$.

Concerning parameters of the fully connected layers, one may consider different layers to capture different information; thus, they should be fine-tuned to a different extent. For this purpose, we employ the so-called discriminative fine-tuning proposed in Ref.~\citenum{Howard2018} for language models. We employ different learning rates for different layers; see \figref{fig:scheme}. We empirically found that our approach performs well with learning rate $\eta_0^{(L)} = 0.01$ of the last layer $L=3$. For lower layers, we fine-tune trainable parameters with learning rates defined by $\eta_0^{(l-1)} = \eta_0^{(l)} / 5$. 

While the discriminative fine-tuning approach is widely employed, there exist other approaches for transfer learning. We experimented with an approach to learn priors for fine-tuning similar to Ref.~\citenum{Shwartz-Ziv2022} but did not find improvements in our experiments. Thus, we did not include the corresponding approach in this work, and more rigorous investigations are postponed to future work. The lacking improvement in the performance on the target task $\mathcal{D}_T$ may be explained by a good alignment of train and test loss surfaces for the investigated task, i.e., the pre-training and fine-tuning tasks are quite similar. This argument is in line with the hypothesis that supervised transfer learning is especially beneficial for closely aligned tasks~\cite{Sun2022}.

All models are trained within the Tensorflow framework~\cite{Abadi2015} on a central processing unit (CPU) node equipped with two Intel Xeon E6252 Gold (Cascade Lake) CPUs.

\section{Results \label{sec:results}}

In this section, we apply the proposed transfer learning approach to two different collections of benchmark data sets, MD17~\cite{Chmiela2017, Schuett2017_2, Chmiela2018, Sauceda2019, Christensen2020b} and ANI~\cite{Smith2018, Smith2019, Smith2020}, and present our results and discussions for a set of experiments designed to assess the quality of fine-tuned interatomic NN potentials. The presented results are obtained by employing the discriminative fine-tuning technique. Thus, the corresponding results for more common approaches, e.g., linear probing, may differ.

\subsection{Molecular dynamics trajectories \label{sec:results_md17}}

One of the main goals of this work is to assess the quality of interatomic NN potentials obtained by the proposed transfer learning approach. Moreover, we are interested in developing models by training only on total energies during fine-tuning but using both energies and atomic forces during pre-training. For this purpose, we employ the MD17 data set originally presented in Refs.~\citenum{Chmiela2017, Schuett2017_2, Chmiela2018} and then revised in Ref.~\citenum{Christensen2020b} to ensure that the respective labels are noise-free. The respective data was sampled from \textit{ab-initio} molecular dynamics (AIMD) simulations. The revised MD17 data is used in this work for pre-training interatomic potentials. It contains 100,000 conformations of each of the ten small organic molecules. The data set includes the respective conformations' structures, energies, and atomic forces. Here, we decided to use only the five molecules aspirin, ethanol, benzene, toluene, and malonaldehyde, since for these molecules, CCSD(T) or CCSD (for aspirin) labels are provided~\cite{Sauceda2019}. For the cutoff radius, we selected a value of $r_\mathrm{c} = 4.0$~\AA{}.

Throughout this work, we will differentiate between four different settings. The conventional setting is training from scratch on coupled-cluster data, i.e., without pre-training on DFT labels. In this setting, energy (E) or atomic forces and energy (F+E) can be used for training. In the transfer learning setting, we use trainable parameters initialized by pre-training interatomic NN potentials on DFT labels for fine-tuning. We also use energy (E) or atomic forces and energy (F+E) for fine-tuning. \figref{fig:md17_learning_curves_with_molecules} compares the learning curves obtained for the four different settings. We used 8192 configurations from the revised MD17 data sets to pre-train our models.

From \figref{fig:md17_learning_curves_with_molecules}, it can be seen that pre-training substantially improves the performance of our potential models. For 950 training configurations and a setting that uses both energy and atomic force labels, we improved the RMS error by a factor of 2--4 and 1.2--2.6 for energy and atomic forces, respectively. For the setting where only energies are used during fine-tuning, we obtained a factor of 3.3--12.3 and 5.0--11.3 for energy and atomic forces, respectively. The largest improvement has been observed for the aspirin molecules, followed by malonaldehyde and toluene. For aspirin, the energy RMS error has been reduced from 4.3~kcal/mol to 0.35~kcal/mol and the atomic force RMS error from 19.5~kcal/mol/\AA{} to 1.7~kcal/mol/\AA{}.
For comparison, the pre-trained model predicts coupled-cluster energies and atomic forces of aspirin molecule with an RMS error of 588.6~kcal/mol and 5.9~kcal/mol/\AA{}, respectively. For other investigated molecules, the respective RMS error is similar.

% E+F: energy, force
% aspirin           0.47/0.23=2.04 1.01/0.83=1.22
% benzene           0.03/0.01=3.00 0.19/0.09=2.11
% ethanol           0.10/0.03=3.33 0.44/0.17=2.59
% malonaldehyde     0.16/0.04=4.00 0.56/0.23=2.43
% toluene           0.18/0.05=3.60 0.55/0.24=2.29

% E: energy, force
% aspirin           4.32/0.35=12.34 19.50/1.73=11.27
% benzene           0.13/0.04= 3.25  1.70/0.34= 5.00
% ethanol           0.60/0.10= 6.00  4.14/0.67= 6.18
% malonaldehyde     0.95/0.11= 8.63  7.27/0.71=10.24
% toluene           0.82/0.10= 8.20  5.23/0.51=10.25

The above results let us make the following statements. First, based on the RMS errors, we see that interatomic NN potentials can efficiently learn atomic forces even though fine-tuning was performed by training only on energies. However, atomic force labels lead to more accurate potentials on par with recent computational results~\cite{Christensen2020b, Pinheiro2021}. In preliminary experiments, we observed that the performance may strongly depend on the electronic structure method used to generate the source data. Thus, the selection of source data should be made with particular attention. Finally, transfer learning leads to more data-efficient models, as we achieve the accuracy of from-scratch-trained models using only a fraction of the data. For example, for aspirin, we need only 128 training structures (F+E) to reach an RMS error of 1.06~kcal/mol/\AA{} in atomic forces by fine-tuning, while 950 training structures are required for training from scratch to reach a similar error (1.01~kcal/mol/\AA{}). For the energy-based fine-tuning on 32 configurations, we reach an accuracy in atomic forces of 6.51~kcal/mol/\AA{}, which is an order of magnitude smaller than the value obtained by training from scratch.

\begin{figure}[t!]
    \centering
    \includegraphics[width=8cm]{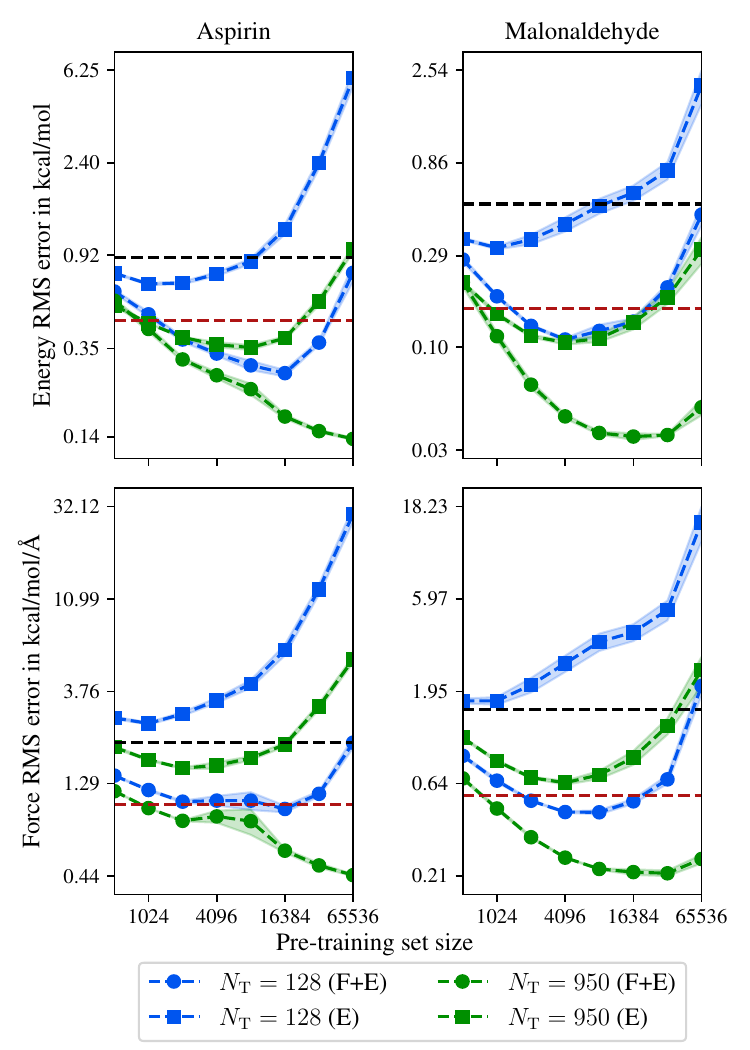}
    \caption{Dependence of the root-mean-square (RMS) errors on the data set size used for pre-training models, fine-tuned with 128 and 950 structures. Shaded areas denote the standard error of the mean evaluated over five independent runs. The horizontal black and red lines denote the final accuracy of models trained from scratch on 128 and 950 coupled-cluster energies and atomic forces, respectively. Simultaneous training on energy and atomic force information (E+F) is compared to energy-based training (E).}
    \label{fig:md17_pretrain_size}
\end{figure}

The data set size used for pre-training may impact the final accuracy, similar to the electronic structure theory used to generate labels. Thus, in \figref{fig:md17_pretrain_size}, we investigate the model's performance dependence on the amount of data used for pre-training when fine-tuned with 128 and 950 structures. We decided to use molecules for which the largest improvement in RMS error has been observed, i.e., aspirin and malonaldehyde.

From \figref{fig:md17_pretrain_size}, we see that the performance of the fine-tuned models can deteriorate when increasing the pre-training data set size past a certain threshold. The threshold appears to depend on the amount of information present in the fine-tuning data set, which depends on the number of fine-tuning structures and whether force labels are available for fine-tuning. For example, when using 128 structures in combination with energy-only (E) fine-tuning, a pre-training set size of 1024 is typically optimal. When increasing the amount of fine-tuning information by either using energy and force (E+F) labels or 950 energy-only (E) fine-tuning structures, optimal pre-training set sizes are typically between 4096 and 16384 structures. When using 950 fine-tuning structures in combination with energy and force (E+F) labels, pre-training set sizes of 32768 (malonaldehyde) or 65535 (aspirin) perform better. For the case of fine-tuning with energy labels only, our empirical results suggest that it is advisable to use pre-training data set sizes of $N_\mathrm{S} \leq 10 N_\mathrm{T}$, where $N_\mathrm{S}$ and $N_\mathrm{T}$ are the pre-training and fine-tuning data set sizes. 

This behavior is unexpected and has not been observed previously~\cite{Smith2019, Kaeser2020, Zheng2021, Kaeser2022, Gardner2022, Zhang2022, Gao2022, Chen2022}. As a possible explanation, we hypothesize that the large number of iterations during pre-training on a large pre-training data set trains the NN to be able to overfit some details of the data more easily, which enables it to overfit the fine-tuning data set more easily. This would suggest that decreasing the number of pre-training epochs when pre-training on very large data sets may allow to circumvent this phenomenon. We leave this as an open question for future work.

As the RMS error is only an abstract measure of the quality of interatomic NN potentials, we run molecular dynamics (MD) simulations, which require a smooth, continuous energy surface to facilitate the numerical integration of the equation of motion. Particularly, we run MD simulations for aspirin molecules in the canonical (NVT) statistical ensemble carried out within the ASE simulation package~\cite{Hjorth2017}. We employ the Langevin thermostat at the temperatures of 100 and 300~K and a time step of 0.5~fs. All MD runs were performed for 110~ps. The sampled MD trajectories are then used to compute velocity-velocity auto-correlation functions and their respective vibrational power spectrum by performing a Fourier transform. The first 10~ps of the dynamics were ignored, and only the remaining 100~ps were used to compute vibrational power spectra.

\begin{figure*}[t!]
    \centering
    \includegraphics[width=16cm]{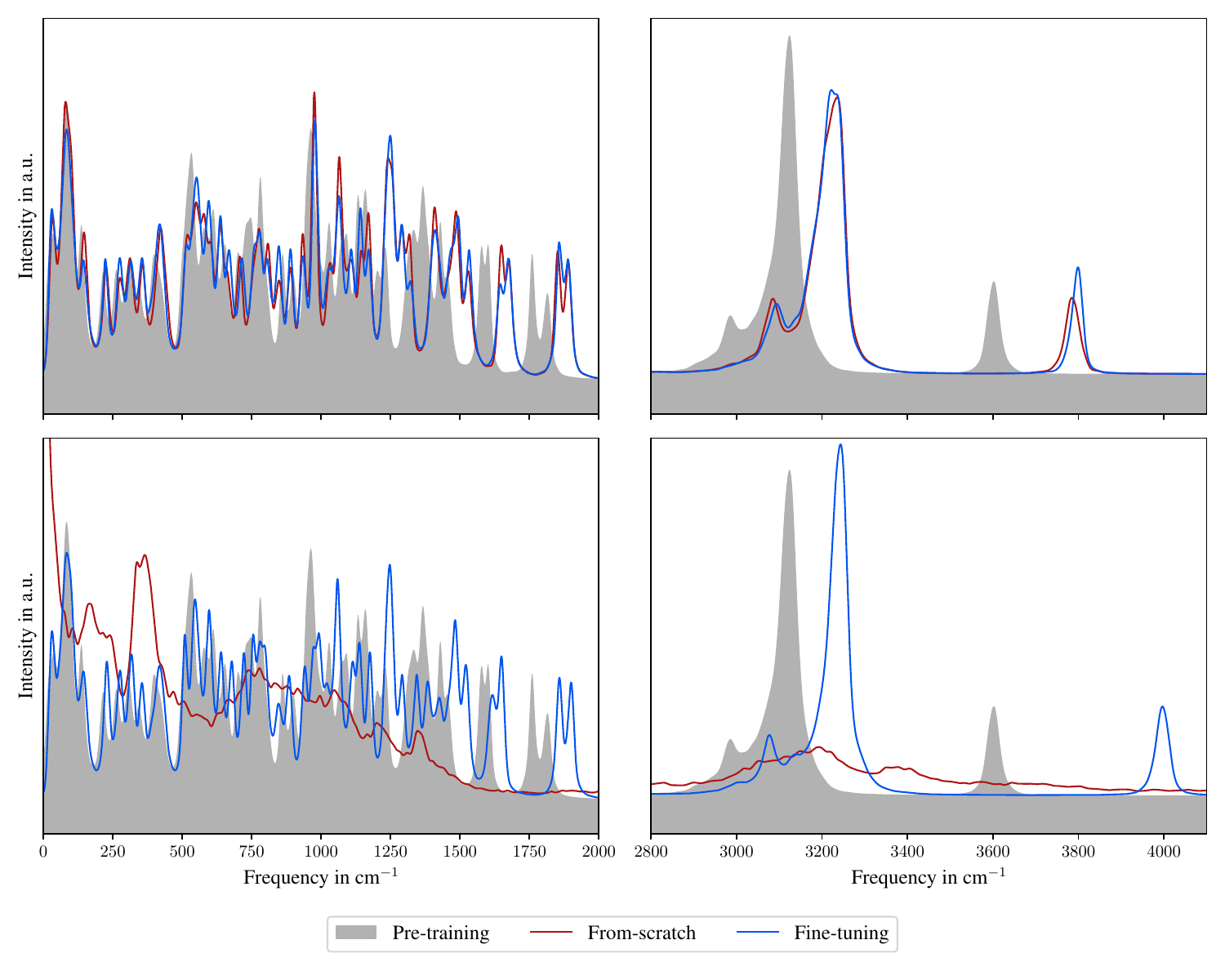}
    \caption{Vibrational power spectrum of the aspirin molecule obtained by computing the Fourier transform of the velocity-velocity auto-correlation function sampled at 300~K. (Top) Comparison of models trained from scratch on 950 and fine-tuned on 128 energy and atomic force labels. (Bottom) Comparison of models trained from scratch and fine-tuned on 950 energy labels only. The characteristic C-H and O-H peaks can be seen around 3200 cm$^{-1}$ and 3800 cm$^{-1}$, respectively.}
    \label{fig:md17_vibrational_dos_300K_0}
\end{figure*}

\figref{fig:md17_vibrational_dos_300K_0} depicts the vibrational power spectrum obtained from MD trajectories sampled at 300~K. First, we assess the quality of potentials obtained by fine-tuning with energies and atomic forces (F+E). From \figref{fig:md17_vibrational_dos_300K_0} (top), we observe that respective power spectra, i.e., obtained from MD simulations run on top of potentials generated by fine-tuning with 128 structures or from scratch with 950 structures, show a similar pattern with negligible differences for O--H and C--H characteristic modes. Thus, fine-tuning with energies and forces leads to qualitatively comparable potentials to those obtained when training from scratch, even though less coupled-cluster data has been used. A comparison of power spectra sampled by models fine-tuned on 128 and 950 structures can be found in the ESI.$^\dag$

The most essential result is, however, the performance of models obtained by fine-tuning on energy labels only, shown in \figref{fig:md17_vibrational_dos_300K_0} (bottom). Here, we observe that models trained from scratch on energies fail and predict a wrong power spectrum. Models fine-tuned on energy labels show improved performance and predict almost all vibrational peaks, which are well aligned with those predicted by models trained from scratch on coupled cluster energy and atomic force data. However, we observe a shift in the frequency compared to our reference coupled-cluster spectrum for the O--H characteristic mode. Also, the intensity of C--H vibrations is sampled slightly worse by models fine-tuned on energy labels than the models trained from scratch or fine-tuned on energies and forces. However, the positions of the corresponding peaks fit well with those obtained with models trained from scratch.

As we show in the ESI$^\dag$, the frequency shift for the O--H characteristic mode for energy-only fine-tuning can be explained by a slightly too steep potential, which otherwise matches the location of the potential minimum of fine-tuning on force and energy labels. We assume that the O--H mode is approximated worse because there is only one O--H bond but many C--H bonds in the aspirin molecule, such that the O--H bond contributes only a smaller part to the total energy. We expect that this issue could be alleviated by either using more data for fine-tuning or generating data with a stronger variance in the O--H vibrations, for example, by suitable active learning or enhanced sampling methods. Note that the total number of scalars in the 128 energy and force labels is 8192, which is considerably larger than the 950 energy labels used for energy-only fine-tuning. More details on the O--H and C--H distance distributions can be found in the ESI.$^\dag$

\subsection{General purpose interatomic potentials \label{sec:results_ani}}

In this section, we assess the proposed transfer learning approach on the ANI-1x and ANI-1ccx data sets in a separate experiment~\cite{Smith2018, Smith2019, Smith2020}. The ANI-1x data set contains configurations, energies, and atomic forces of 4,956,005 molecules generated through an active learning approach~\cite{Smith2018}. The respective labels are obtained from density functional calculations. The ANI-1ccx data set contains configurations and energies of 489,571 molecules~\cite{Smith2019}, while the corresponding labels are computed at the CCSD(T)/CBS level of theory. For the cutoff radius of our interatomic NN potentials, we selected a value of $r_\mathrm{c} = 5.0$~\AA{}.

\begin{table}[t!]
\centering
\caption{\label{tab:table_comp6} Mean absolute (MAE) and root-mean-square (RMSE) errors in predicted energies/forces for the COMP6 benchmark data set~\cite{Smith2018}. Total energies are given in kcal/mol, while forces are given in kcal/mol/\AA{}.}
\begin{tabular}{ccccccc}
\hline
Training set size& & \multicolumn{2}{c}{524,288} &\multicolumn{2}{c}{4,194,304} \\
& & MAE & RMSE & MAE & RMSE\\
\hline
\multirow{2}{*}{ANI-MD} 	    & energy & 3.83 & 7.06 & 3.48 & 6.41 \\
 	                            & force  & 1.43 & 2.57 & 1.32 & 2.47 \\
\multirow{2}{*}{DrugBank} 	    & energy & 2.78 & 4.21 & 2.57 & 4.18 \\
 	                            & force  & 1.69 & 2.82 & 2.61 & 1.55 \\
\multirow{2}{*}{GDB07to09} 	    & energy & 1.22 & 1.61 & 1.09 & 1.44 \\
 	                            & force  & 1.41 & 2.24 & 1.24 & 1.95 \\
\multirow{2}{*}{GDB10to13} 	    & energy & 2.29 & 3.04 & 2.08 & 2.76 \\
 	                            & force  & 2.25 & 3.66 & 2.02 & 3.26 \\
\multirow{2}{*}{S66x8} 	        & energy & 2.95 & 4.04 & 2.71 & 3.78 \\
 	                            & force  & 0.93 & 1.67 & 0.86 & 1.57 \\
\multirow{2}{*}{Tripeptides}    & energy & 3.06 & 4.35 & 2.73 & 3.65 \\
                                & force  & 1.48 & 4.46 & 1.32 & 2.61 \\
\hline
\multirow{2}{*}{COMP6}          & energy & 2.03 & 3.02 & 1.83 & 2.79 \\
                                & force  & 1.85 & 3.11 & 1.65 & 2.74 \\
\hline
\end{tabular}
\end{table}

The training of interatomic NN potentials on the ANI-1x data set is challenging. Thus, we discuss the performance of models pre-trained on the ANI-1x data set before fine-tuning experiments. To assess the accuracy of pre-trained models, we employ the COMP6 benchmark data set~\cite{Smith2018}. We provide the results for data sets included in COMP6 and for COMP6 as a whole. The individual results for training set sizes of 524,288 and 4,194,304 are shown in \tabref{tab:table_comp6}. We compare our model to the performance of the well-established ANI and equivariant message-passing NewtonNet models~\cite{Smith2017, Haghighatlari2022}. Trained on 4,956,005 molecules, ANI achieves an MAE of 1.61 kcal/mol and 2.70~kcal/mol/\AA{} in predicted energies and atomic forces, respectively~\cite{Haghighatlari2022}.

For the equivariant NewtonNet models trained on 495,600 molecules, an MAE of 1.45~kcal/mol and 1.79~kcal/mol/\AA{} for the energies and atomic forces have been reported~\cite{Haghighatlari2022}. In our experiments, we have found that GM-NN models trained on 524,288 molecules perform close to the equivariant message-passing NewtonNet model and achieve an MAE of 2.03~kcal/mol and 1.85~kcal/mol/\AA{} for energies and atomic forces, respectively. The comparable performance of both models can be attributed to the similarity in the underlying ideas of our and equivariant message-passing frameworks. Both approaches apply equivariant transformations to the input coordinates and subsequently build features invariant to rotations.

\begin{figure}[t!]
    \centering
    \includegraphics[width=8cm]{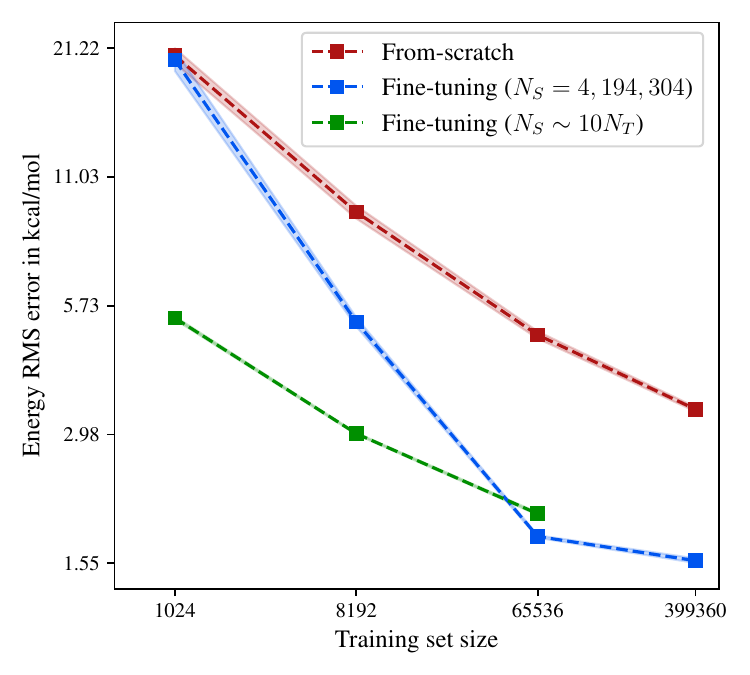}
    \caption{Learning curves for the ANI-1ccx data set~\cite{Smith2019, Smith2020}. The root-mean-square (RMS) errors of total energies are plotted against the training set size. Shaded areas denote the standard error of the mean evaluated over five independent runs. Models trained directly on the coupled-cluster data are referred to as ``from scratch''. Models obtained by employing the fine-tuning approach to transfer from density functional-based pre-trained models are denoted by ``fine-tuning''. All models were trained on energy labels only.}
    \label{fig:ani_learning_curves}
\end{figure}

Since the ANI-1ccx data set does not contain force labels, only the energy-training setting (E) can be used for both from-scratch training and fine-tuning.  \figref{fig:ani_learning_curves} compares the learning curves obtained for the two different settings. We used models pre-trained on 4,194,304 molecules from the ANI-1x data set to initialize trainable parameters. From \figref{fig:ani_learning_curves}, one can observe, similar to \secref{sec:results_md17}, that pre-training of interatomic NN potentials substantially improves their sample efficiency. Using 65,536 molecules for fine-tuning, we obtained an energy RMS error of 1.77~kcal/mol, while for training from scratch on 399,360 molecules, an error of 3.39~kcal/mol could be achieved. By increasing the training data set size for fine-tuned models to 399,360, we get an energy RMS error of 1.57~kcal/mol. Finally, we observed the same tendency concerning the fine-tuning and pre-training data set sizes compared to \secref{sec:results_md17}. \figref{fig:ani_learning_curves} shows that for small fine-tuning data set sizes, i.e., $N_\mathrm{T}<65,536$ for ANI-1ccx, the pre-training data set sizes should not exceed $\sim 10N_\mathrm{T}$. However, for larger fine-tuning data set sizes, the performance seems insensitive to the pre-training data set size.

\begin{figure}[t!]
    \centering
    \includegraphics[width=8cm]{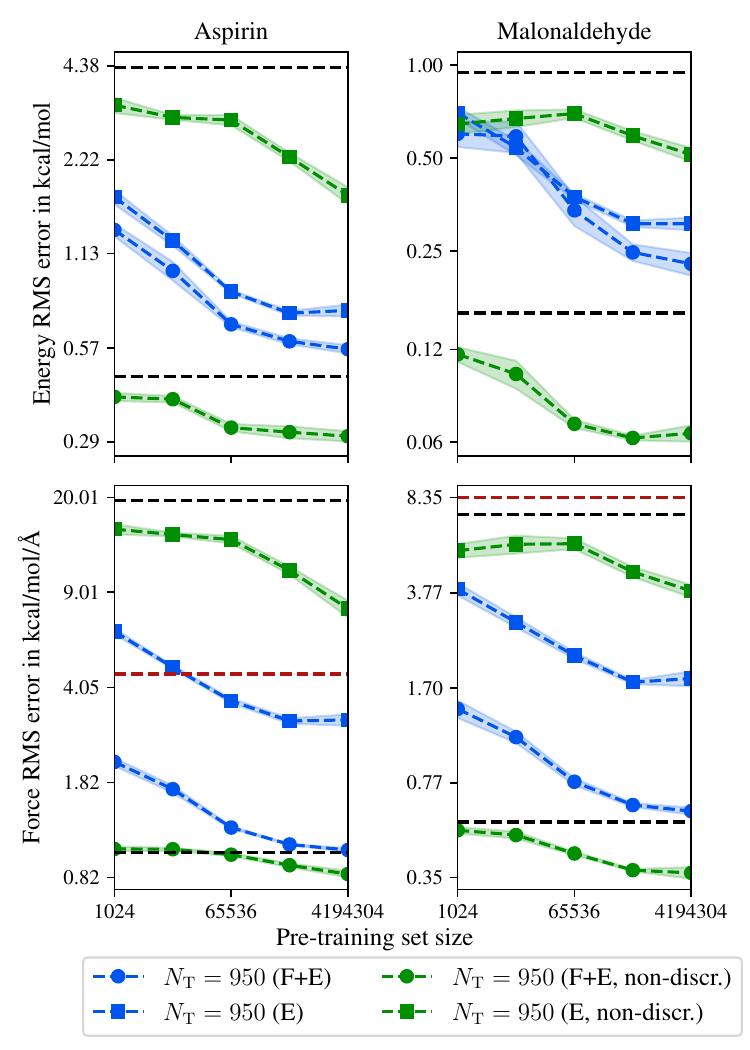}
    \caption{Dependence of the root-mean-square (RMS) errors on the ANI-1x data set size used for pre-training models, fine-tuned with 950 structures from MD17. Shaded areas denote the standard error of the mean evaluated over five independent runs. The green lines correspond to non-discriminative fine-tuning (non-discr.) with the default learning rates for training from scratch: $\eta_0^{\left(l\right)} = \eta_0^\mathrm{base} = 0.03$, $\eta_0^\mathrm{embed} = 0.02$, $\eta_0^\mathrm{shift}=0.05$, and $\eta_0^\mathrm{scale}=0.001$. The horizontal black lines denote the final accuracy of models trained from scratch on 950 coupled-cluster energies (higher values) or energies and atomic forces (lower values). The horizontal red line denotes the performance of models pre-trained on 4,194,304 structures from ANI-1x. Simultaneous training on energy and atomic force information (E+F) is compared to energy-based training (E).}
    \label{fig:ani_md17_pretrain_size}
\end{figure}

To assess the dependence of the final model's performance on the pre-training data set size more rigorously, we fine-tuned models pre-trained on the ANI-1x data set using 950 structures from MD17 at the CCSD level. \figref{fig:ani_md17_pretrain_size} shows that the performance of fine-tuned models is improved with increasing pre-training data set size. This behavior is observed for training on energies and atomic forces as well as on energies only, in contrast to results from \secref{sec:results_md17}, although the maximal pre-training set sizes are much larger here. Our observation suggests that for substantially different pre-training data sets, the optimal amount of pre-training data should not generally be estimated by the number of structures but rather by the relation of the pre-training error to the achievable error of from-scratch training. The models pre-trained on 4,194,304 structures from the ANI-1x data set achieve a force RMS error of 2.74~kcal/mol/\AA{}, which is less than an order of magnitude smaller than the achievable from-scratch error for 950 energy-only aspirin structures, while the respective pre-training error on 8192 aspirin structures from the MD17 data set is about 0.43~kcal/mol/\AA{}. Hence, we do not observe overfitting effects even with a large pre-training set here.

Aside from studying the pre-training data set size dependence, we observed that ANI general purpose potentials can be used to improve the performance by fine-tuning on energies of small organic molecules. For aspirin, the RMS error in atomic forces has been reduced from 19.50 to 3.08~kcal/mol/\AA{} by employing transfer learning. However, we did not observe any improvement when fine-tuning on energies and atomic forces compared to training from scratch, which already achieves a lower force error than the pre-trained model. As an explanation, we investigate the hypothesis that because discriminative fine-tuning hinders an adaptation of earlier NN layers and, in particular, the Gaussian moments descriptor, the fine-tuning error cannot improve much on the pre-training error, which can be seen as underfitting. Indeed, our results in \figref{fig:ani_md17_pretrain_size} show that in this case, by using non-discriminative fine-tuning with the learning rates for from-scratch training, we can outperform both discriminative fine-tuning and from-scratch training. On the other hand, for fine-tuning with energies only, where from-scratch training performs poorly, discriminative fine-tuning performs better than non-discriminative fine-tuning.

\begin{figure*}[t!]
    \centering
    \includegraphics[width=16cm]{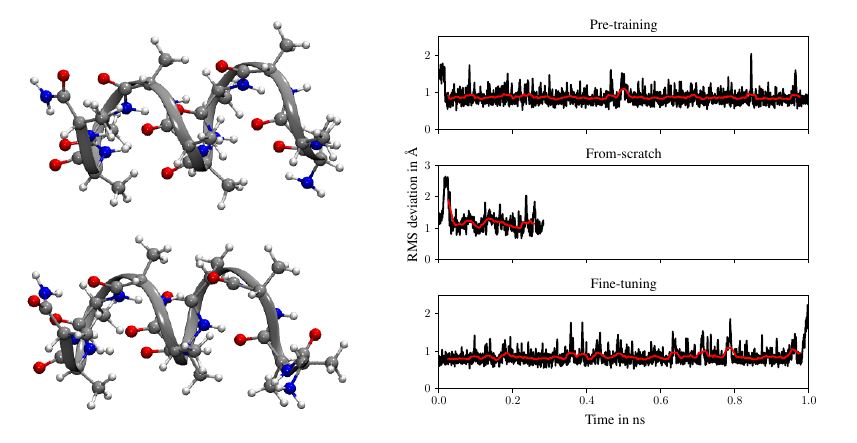}
    \caption{Root-mean-square (RMS) deviation of 104-atom deca-alanine (Ala$_{10}$) in the gas phase with respect to the initial helical structure evaluated for the 1~ns-long MD simulation. Atomic forces used to run the respective MD simulations are obtained from an ensemble of five interatomic NN potentials (top) trained on 4,194,304 molecules from ANI-1x data sets and (bottom) subsequently fine-tuned on 399,360 molecules from the ANI-1ccx data set, or (middle) trained from scratch on 399,360 molecules from the ANI-1ccx data set. The red line indicates the respective running average with a window size of 25~ps. The (top) initial Ala$_{10}$ structure as well as (bottom) an observed conformation of Ala$_{10}$ are shown as an inset for models obtained by fine-tuning.}
    \label{fig:tikz-deca-alanine-300K}
\end{figure*}

Assessing the quality of atomic forces predicted by models fine-tuned on ANI-1ccx is hardly possible as computing force labels at the CCSD(T)/CBS level of theory is infeasible on standard compute nodes. Thus, we assess the overall quality of developed potentials by running molecular dynamics simulations by employing forces obtained by an ensemble of five interatomic potential models. As pre-training and fine-tuning models, we use models trained on 4,194,304 molecules from ANI-1x data sets and subsequently fine-tuned on 399,360 molecules from the ANI-1ccx data set, respectively. As models trained from scratch, we use models trained directly on 399,360 molecules from the ANI-1ccx data set. We run MD simulations at 300~K for 104-atom deca-alanine (Ala$_{10}$), frequently used to study protein folding dynamics~\cite{Hazel2014}. Deca-alanine is not part of ANI-1x and ANI-1ccx data sets. We investigate molecular dynamics trajectories of Ala$_{10}$ in the gas phase; see \figref{fig:tikz-deca-alanine-300K}. For this purpose, we run MD simulations in the canonical (NVT) statistical ensemble carried out within the ASE simulation package~\cite{Hjorth2017}. We employ the Langevin thermostat at a temperature of 300~K and a time step of 0.5~fs. We run the MD simulations for 1.1~ns each. The first 100~ps are used for the equilibration and excluded from the analysis.

One of the main observations is that for models trained from scratch on 399,360 molecules from the ANI-1ccx data set, it was impossible to run a stable MD simulation with Ala$_{10}$ in the gas phase for more than $\sim 350$~ps. In contrast, models fine-tuned on 399,360 molecules from the respective data set led to a stable MD simulation run over 1.1~ns. \figref{fig:tikz-deca-alanine-300K} shows the RMS deviation of Ala$_{10}$ in the gas phase with respect to the initial configuration for pre-trained and fine-tuned interatomic NN models, as well as models trained from scratch. The helical structure of Ala$_{10}$ is preserved for both models and only minor conformational changes have been observed; see \figref{fig:tikz-deca-alanine-300K}. Particularly, only rotations of a terminal alanine residue around the corresponding C--C bond have been observed.

\begin{figure*}[t!]
    \centering
    \includegraphics[width=16cm]{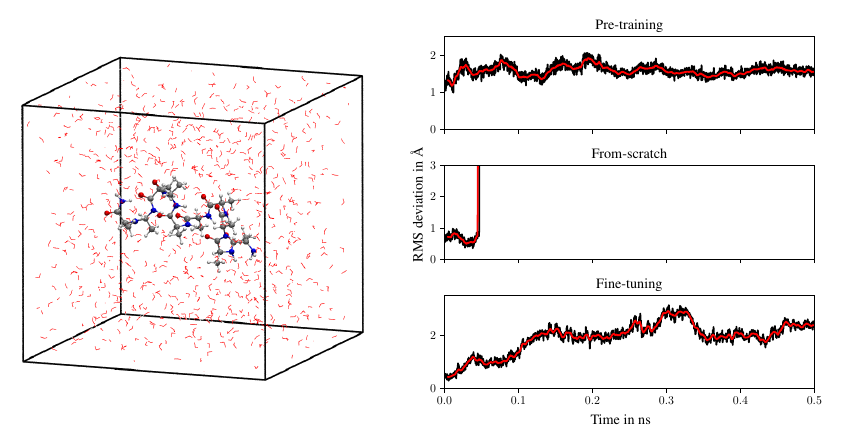}
    \caption{Root-mean-square (RMS) deviation of deca-alanine (Ala$_{10}$) in water (2384 atoms) with respect to the initial helical structure evaluated for the 0.5~ns-long MD simulation. Atomic forces used to run the respective MD simulations are obtained from an ensemble of five interatomic NN potentials (top) trained on 4,194,304 molecules from ANI-1x data sets and (bottom) subsequently fine-tuned on 399,360 molecules from the ANI-1ccx data set, or (middle) trained from scratch on 399,360 molecules from the ANI-1ccx data set. The red line indicates the respective running average with a window size of 5~ps. An example structure is shown as an inset.}
    \label{fig:tikz-deca-alanine-solvate-300K}
\end{figure*}

Using the results from Ref.~\citenum{Zaverkin2022c}, i.e., that local models trained on cluster data can be used to predict periodic bulk structures, we ran MD simulations for the Ala$_{10}$ molecule in water; see \figref{fig:tikz-deca-alanine-solvate-300K}. For this purpose, we pre-equilibrated the atomic system by running 10~ps-long MD simulations employing Langevin and Nos{\'e}--Hoover thermostats at 300~K and a time step of 0.5~fs sequentially. Then, for production simulations, we use the isobaric-isothermal form of the Nos{\'e}--Hoover dynamics~\cite{Melchionna1993, Melchionna2000}, as implemented in ASE~\cite{Hjorth2017}, at $T = 300$~K and $p = 1$~bar. The respective MD simulations are run over 510~ps, while the first 10~ps are reserved for equilibration. A time step of 0.5~fs has been used to integrate the equations of motion, while the characteristic time scales of the thermostat and barostat were set to 1~ps each. The simulation box was allowed to change independently along the three Cartesian axes, $x$, $y$, and $z$. However, the angle between axes has been fixed to 90 degrees.

To run the dynamics of Ala$_{10}$ in the bulk water (2384 atoms), we employed the same ensemble models used to run MD simulations for the Ala$_{10}$ molecule in the gas phase. The RMS deviation of Ala$_{10}$ with respect to the initial configuration is shown in \figref{fig:tikz-deca-alanine-solvate-300K}. For models pre-trained on the ANI-1x data set, we observed that the protein stays in its helical state over the course of the simulation. In contrast, we observed low-helical states of deca-alanine states when using atomic forces provided by models fine-tuned on ANI-1ccx, on par with the recent computational results~\cite{Hazel2014}. When employing models trained from scratch on the ANI-1ccx data set, it was impossible to run a stable MD simulation for more than $\sim 40$~ps. Note that we are not going to perform a detailed analysis of the protein folding in this work and are aiming to show the advantage of using transfer learning approaches for developing interatomic NN potentials. Moreover, a thorough assessment of the fine-tuning data set is required prior to running real-world applications, as missing configurations may lead to locally inaccurate potentials as shown in \figref{fig:md17_vibrational_dos_300K_0} and discussed in \secref{sec:results_md17}.

\section{Discussion and conclusion \label{sec:discussion_conclusion}}

This work investigated a transfer learning approach to modeling chemically accurate interatomic interactions by neural networks. We initialized trainable parameters from models pre-trained on labels obtained from density functional calculations and then fine-tuned the respective parameters using, e.g., coupled-cluster labels. In our approach, we fixed the trainable parameters of the local atomic representation but re-initialized the trainable scale and shift parameters of atomic energies. Moreover, we employed discriminative fine-tuning~\cite{Howard2018} for fully connected layers to ensure that different layers are optimized to a different extent, as they may contain different information that requires different amounts of adjustment. Note that the results may differ if different transfer learning approaches, e.g., linear probing, or different hyper-parameters, are used.

The proposed transfer learning approach has been tested on two different benchmark data sets, MD17~\cite{Chmiela2017, Schuett2017_2, Chmiela2018, Sauceda2019, Christensen2020b} and ANI~\cite{Smith2018, Smith2019, Smith2020}. Here, particular attention is drawn to the overall applicability of transfer learning approaches to modeling interatomic interactions and their sample efficiency compared to models trained from scratch on the respective labels. Moreover, the MD17 data set provides coupled-cluster labels for total energies and atomic forces~\cite{Sauceda2019}. It thus provides the means to investigate the performance of force prediction for models obtained by fine-tuning on only energy labels more thoroughly than ANI.

We have found that transfer learning approaches lead to more sample-efficient models, i.e., models requiring fewer computationally expensive \textit{ab-initio} labels compared to training interatomic NN models from scratch. For example, we required about seven times fewer energy and atomic force labels at the fine-tuning level to obtain the same accuracy on aspirin molecules compared to the models trained from scratch. Moreover, for the setting where only total energies have been used for fine-tuning, we achieved a force RMS error of 6.51~kcal/mol/\AA{}, which is a third of the error when training from scratch. In addition, the models fine-tuned on total energies of 950 aspirin molecules achieved a force RMS error of 1.73~kcal/mol/\AA{}, while the models trained from scratch on energy and atomic force labels we obtained an error of 1.01~kcal/mol/\AA{}. For the errors in predicted total energies, similar results have been obtained.

Similar to MD17 experiments, we have observed an improved data efficiency for models obtained by fine-tuning with coupled-cluster energies from the ANI-1ccx data set~\cite{Smith2019, Smith2020}, which covers diverse molecular compositions and conformations. However, in this case, we could not assess the accuracy of predicted forces, as they were unavailable in the data set. For predicted energies, we could reduce the required training set size by a factor of ten compared to training from scratch. This is a considerable improvement considering the high computational cost of CCSD(T)/CBS labels.

Besides the performance of fine-tuned models, we investigated the performance of GM-NN models, pre-trained on the ANI-1x data set, on the COMP6 data set~\cite{Smith2018, Smith2020}. We have found that our interatomic NN potentials on diverse data sets have data efficiency and accuracy on par with equivariant message-passing architectures, e.g., NewtonNet~\cite{Haghighatlari2022}. For GM-NN trained on 524,288 molecules, we obtained MAEs of 2.03~kcal/mol and 1.85~kcal/mol/\AA{} for energies and atomic forces, respectively.  In comparison, for NewtonNet trained on 495,600 molecules, an MAE of 1.45~kcal/mol and 1.79~kcal/mol/\AA{} for the corresponding properties has been reported.

Besides the improved data efficiency of our models employing transfer learning, we studied the dependence of their performance on the pre-training data set size. For the MD17 data set, we observed that models fine-tuned from parameters obtained by pre-training on too large data sets performed worse than using parameters initialized by pre-training on smaller data sets. Notably, when fine-tuning with energies only, we have found that pre-training data set size should maximally exceed the fine-tuning data set size tenfold. A similar observation has been made for the ANI data sets. However, here, we observed that for extensive fine-tuning data sets, e.g., sized $>60,000$ molecules, the dependence of the performance on the pre-training data set size vanishes. Additionally, it appears that very large pre-training set sizes can be beneficial as long as the achieved RMS error is not too small. In the case where the pre-training error is larger than the error for training from scratch, we observed that discriminative fine-tuning can lead to underfitting, which can be resolved by using larger learning rates for earlier layers during fine-tuning as well.

The excellent performance of our fine-tuned models allowed us to assess the performance of developed interatomic NN potentials during a molecular dynamics simulation. For this purpose, we have used the largest molecule from the MD17 data set---the aspirin molecule. In these experiments, we observed that models obtained by fine-tuning with energy labels lead to more stable trajectories than the corresponding models trained from scratch. However, training from scratch or fine-tuning with energy and force labels both led to good results. Note that we employed models fine-tuned on 128 energy and force labels from coupled-cluster calculations, while models trained from scratch used the respective labels of 950 structures. From the obtained trajectories, we could compute molecular vibrational spectra. While models trained from scratch on energy labels failed to predict useful vibration spectra, other models trained from scratch on energies and atomic forces or fine-tuned on energies or energies and atomic forces could do this reasonably well.

We obtained almost indistinguishable spectra by using models trained from scratch or fine-tuning with energy and force labels. By running molecular dynamics with forces provided by models fine-tuned on solely energy labels we could reproduce most of the vibrational spectra well, except for a moderate difference in intensity for the C--H peak and a shift in position for the O--H peak. However, we expect that this could be resolved by using more or more carefully selected fine-tuning data. In any case, the highlighted shortage of transfer learned potentials should not be considered as the limitation of the method but of the underlying data set, which has to be designed carefully to sample all relevant vibrational degrees of freedom. Finally, fine-tuning on energy and atomic force labels will consistently outperform fine-tuning on energy labels only. Thus, the respective setting is advised solely for systems for which atomic forces are inaccessible at the desired level of theory.

To investigate the generalization abilities of our potentials fine-tuned on ANI-1ccx data, we ran molecular dynamics simulation with a larger molecule, deca-alanine (Ala$_{10}$) in the gas phase (104 atoms) and water (2384 atoms). While models trained from scratch on 399,360 molecules from the ANI-1ccx data set could not be used to run stable dynamics for longer than $\sim 350$~ps, models obtained by fine-tuning on 399,360 molecules could do so. We investigated the RMS deviation of Ala$_{10}$ with respect to its initial configuration in the gas phase and water. We found that it stays in its helical state over the course of the simulation or changes its state to a low-helical one, on par with recent computational results~\cite{Hazel2014}. However, a detailed analysis of the protein folding dynamics is out of the scope of this paper. With these experiments, we aimed to show the potential capability of fine-tuned interatomic NN potentials to investigate bio-molecular systems while preserving the chemical accuracy of the reference method.

In summary, this work proposes an alternative transfer learning approach for fine-tuning interatomic NN potentials with computationally expensive \textit{ab-initio} labels. We demonstrate that models obtained by fine-tuning on energy labels only can be used for large-scale simulations and provide a means of investigating complex biological matter. However, particular attention should be drawn to designing appropriate pre-training and fine-tuning data sets, as missing atomic force labels may lead to losing essential information for developing reliable interatomic potentials. In this respect, it might be interesting to consider the generation of pre-training data sets using NN potentials instead of DFT~\cite{Gardner2022}, for example using our pre-trained ANI model.

\section*{Author Contributions}
VZ: Conceptualization, Methodology, Software, Investigation, Writing -- Original Draft; DH: Conceptualization, Methodology, Writing -- Review \& Editing; LB: Investigation, Writing -- Review \& Editing; JK: Supervision, Funding acquisition, Writing -- Review \& Editing.

\section*{Conflicts of interest}
There are no conflicts to declare.

\section*{Acknowledgements}

Funded by Deutsche Forschungsgemeinschaft (DFG, German Research Foundation) under Germany's Excellence Strategy - EXC 2075 - 390740016 and the Ministry of Science, Research and the Arts Baden-W{\"u}rttemberg in the Artificial Intelligence Software Academy (AISA). We acknowledge the support by the Stuttgart Center for Simulation Science (SimTech). The authors acknowledge support by the state of Baden-W{\"u}rttemberg through bwHPC and the German Research Foundation (DFG) through grant no INST 40/575-1 FUGG (JUSTUS 2 cluster). The authors thank the International Max Planck Research School for Intelligent Systems (IMPRS-IS) for supporting David Holzm{\"u}ller.

\bibliography{main}

%%%%%%%%%% Merge with supplemental materials %%%%%%%%%%
\clearpage

\begin{center}
\textbf{\Large Supplementary Information \\ Transfer learning for chemically accurate interatomic neural network potentials}
\end{center}

\setcounter{section}{0}
\setcounter{equation}{0}
\setcounter{figure}{0}
\setcounter{table}{0}
\setcounter{page}{1}
\makeatletter
\renewcommand{\theequation}{S\arabic{equation}}
\renewcommand{\thefigure}{S\arabic{figure}}
\renewcommand{\thesection}{S-\Roman{section}}

\section{Molecular dynamics trajectories}

\begin{figure*}[h!]
    \centering
    \includegraphics[width=16cm]{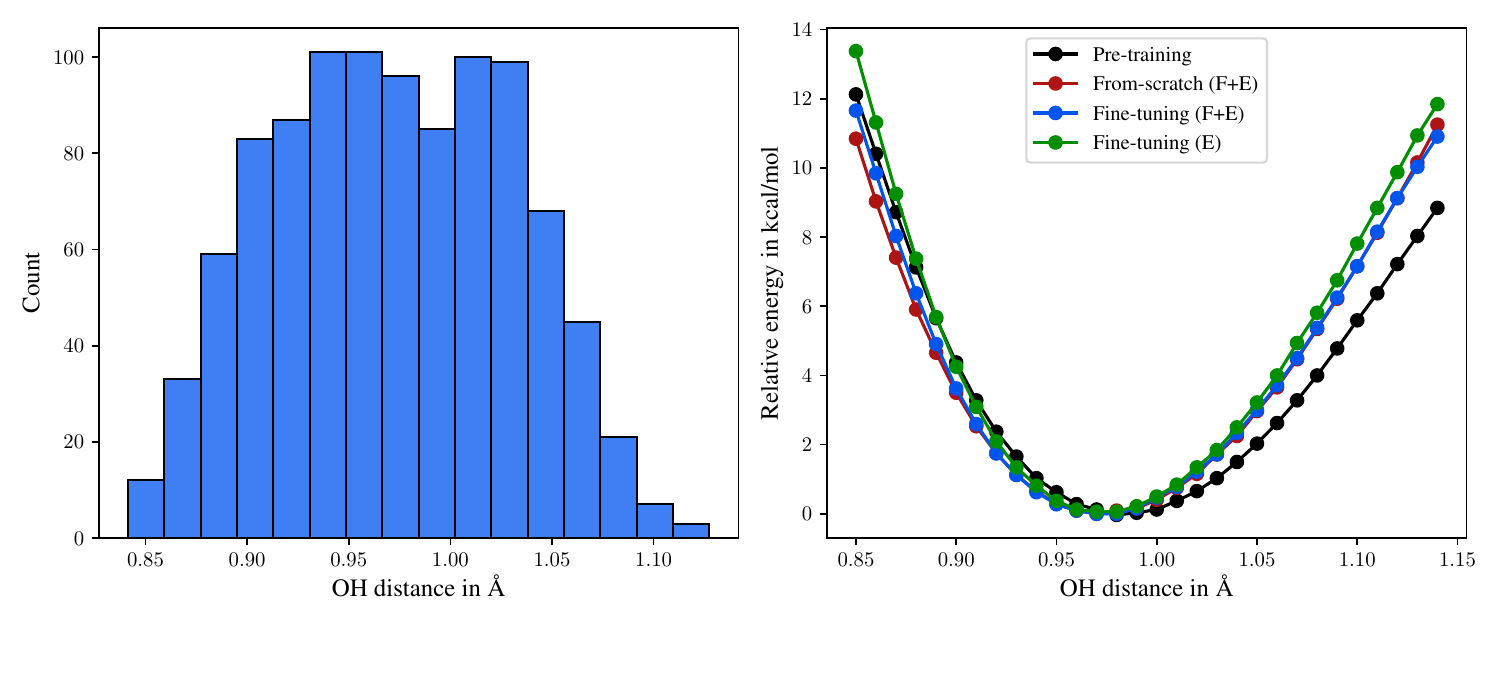}
    \caption{(Left) Distribution of O--H distances in the coupled-cluster aspirin data set. (Right) Relative energy dependence on the O--H distance for four interatomic potentials.}
    \label{fig:oh_distances}
\end{figure*}

\begin{figure*}[h!]
    \centering
    \includegraphics[width=16cm]{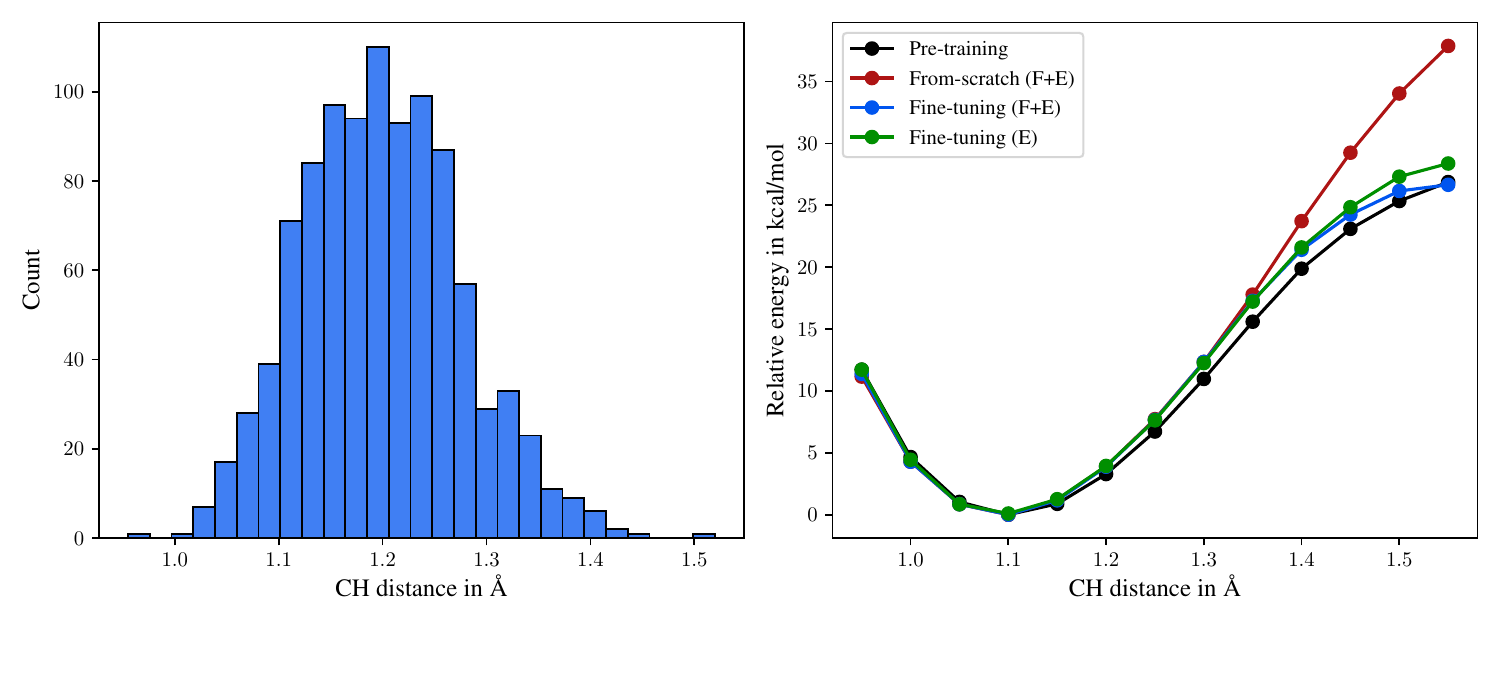}
    \caption{(Left) Distribution of C--H distances in the coupled-cluster aspirin data set. (Right) Relative energy dependence on the C--H distance for four interatomic potentials.}
    \label{fig:ch_distances}
\end{figure*}

Here, additional results for the aspirin molecule from the MD17 data set~\cite{Chmiela2017, Schuett2017_2, Chmiela2018, Sauceda2019, Christensen2020b} are presented. Particularly, we study O--H and C--H distance distributions of the coupled-cluster aspirin data set~\cite{Sauceda2019} to explain the observed deviations for C--H and O--H characteristic modes. \Figref{fig:oh_distances} and \Figref{fig:ch_distances} respresent the correspoding results for C--H and O--H distances. From \figref{fig:oh_distances} (left), we see that the respective O--H distances vary between 0.84 and 1.13~\AA{}. However, the respective counts decrease by approaching the boundary values, indicating a somewhat worse sampling of high-energy regions.

From \figref{fig:oh_distances} (right), we observe a steeper potential energy surface for the model fine-tuned on energy values compared to the model trained on energies and forces of 950 configurations from scratch. The model fine-tuned on energy and forces of 128 configurations matches the latter well. Fitting the respective one-dimensional potential energy surfaces by a squared function, we could estimate vibrational frequencies of 3532, 3558, 3589, 3793~$cm^{-1}$ for the pre-trained model, the model trained from scratch and models fine-tuned on energy and force or energy labels, respectively. Note that the calculated values may strongly depend on the fitting procedure and serve only as a rough estimate. The obtained results support our observations in the main manuscript. Moreover, these results support the necessity of a thorough data set generation when fine-tuning with energy values only, e.g., better sampling of high-energy regions or augmenting data with more configurations and respective energy labels.

In contrast, from \figref{fig:ch_distances}, we observe that the C--H distances have been slightly better sampled than O--H distances. In addition, the aspirin molecule has more C--H bonds than O--H bonds. This fact could explain why only minor deviations of fine-tuned models from those trained from scratch can be seen. The results in \figref{fig:ch_distances} match our observation that the C--H characteristic mode is predicted better than the O--H characteristic mode.

\begin{figure*}[t!]
    \centering
    \includegraphics[width=16cm]{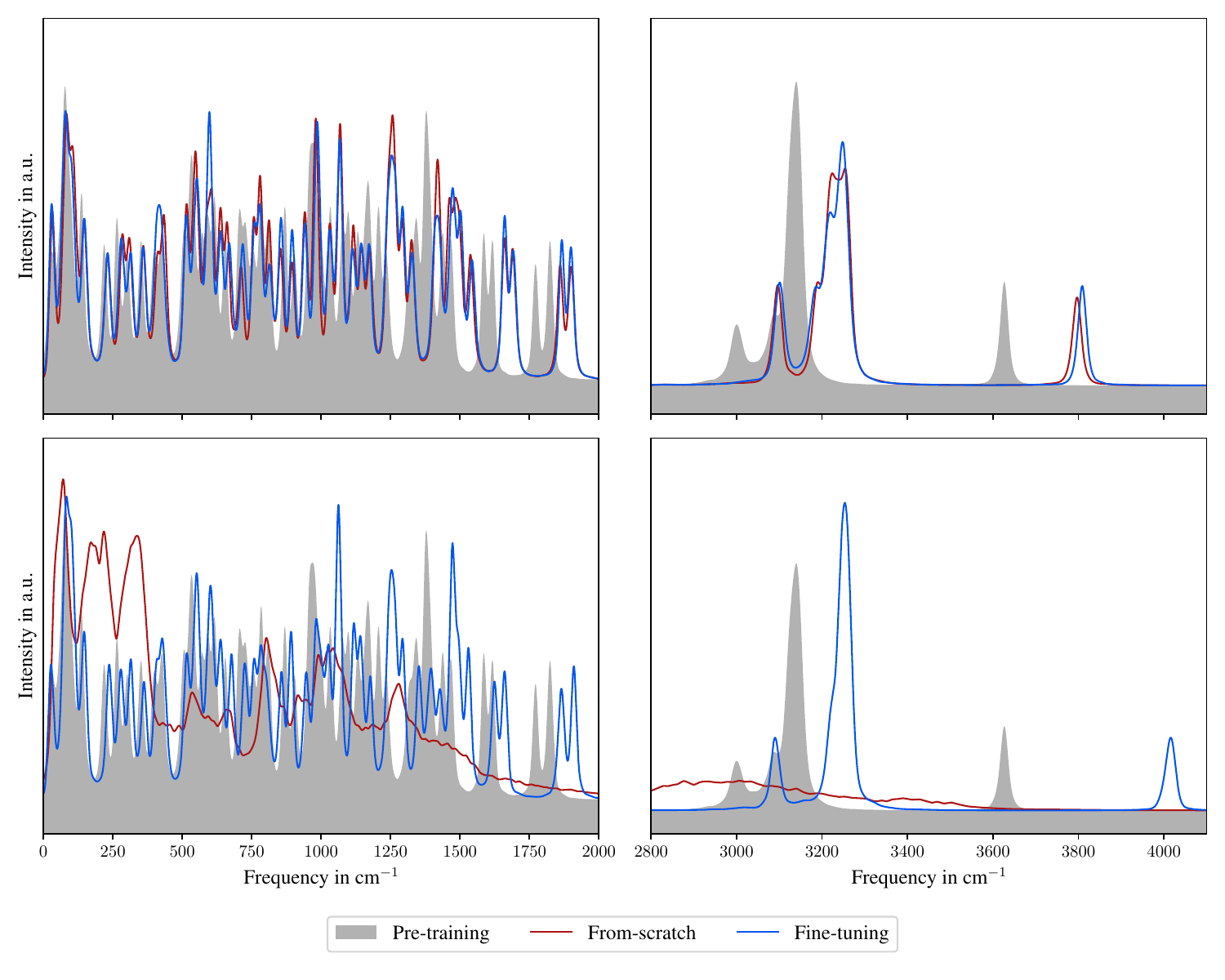}
    \caption{Vibrational power spectrum of the aspirin molecule obtained by computing the Fourier transform of the velocity-velocity auto-correlation function sampled at 100~K. (Top) Comparison of models trained from scratch on 950 and fine-tuned on 128 energy and atomic force labels. (Bottom) Comparison of models trained from scratch and fine-tuned on 950 energy labels only. The characteristic C-H and O-H peaks can be seen around 3200 cm$^{-1}$ and 3800 cm$^{-1}$, respectively.}
    \label{fig:md17_vibrational_dos_100K_0}
\end{figure*}

\begin{figure*}[t!]
    \centering
    \includegraphics[width=16cm]{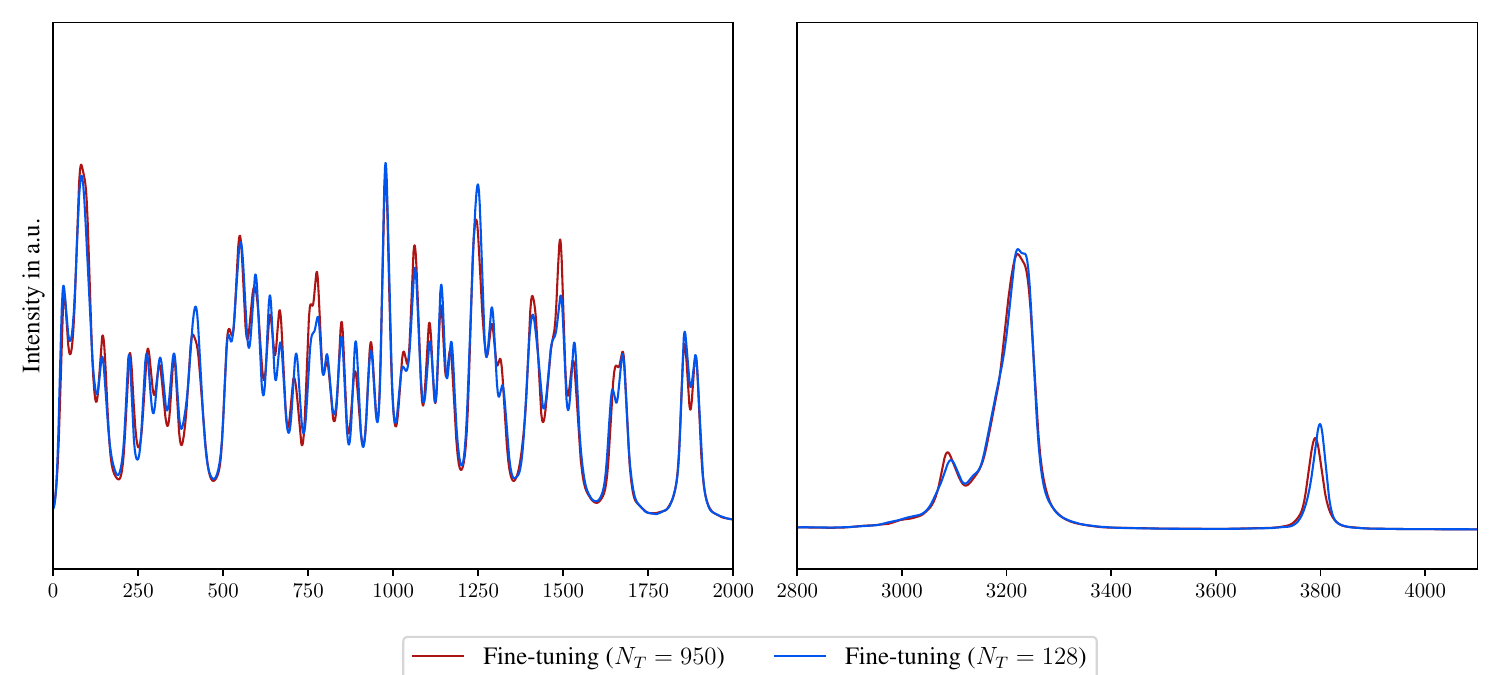}
    \caption{Vibrational power spectrum of the aspirin molecule obtained by computing the Fourier transform of the velocity-velocity auto-correlation function sampled at 300~K. Comparison of models fine-tuned on 128 and 950 energy and atomic force labels. The characteristic C-H and O-H peaks can be seen around 3200 cm$^{-1}$ and 3800 cm$^{-1}$, respectively.}
    \label{fig:md17_vibrational_dos_300K_950_vs_128}
\end{figure*}

\begin{figure*}[t!]
    \centering
    \includegraphics[width=16cm]{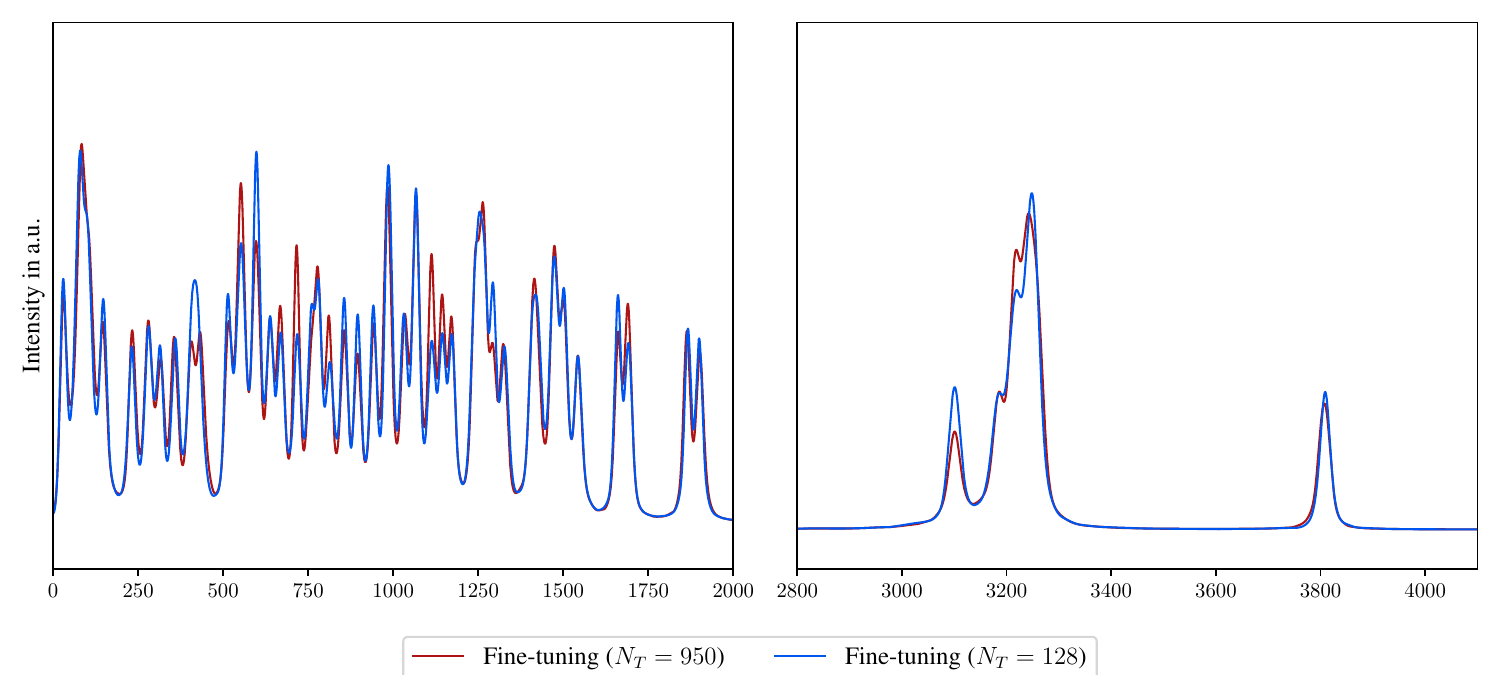}
    \caption{Vibrational power spectrum of the aspirin molecule obtained by computing the Fourier transform of the velocity-velocity auto-correlation function sampled at 100~K. Comparison of models fine-tuned on 128 and 950 energy and atomic force labels. The characteristic C-H and O-H peaks can be seen around 3200 cm$^{-1}$ and 3800 cm$^{-1}$, respectively.}
    \label{fig:md17_vibrational_dos_100K_950_vs_128}
\end{figure*}

\Figref{fig:md17_vibrational_dos_100K_0} shows the vibrational power spectrum of the aspirin molecule obtained by computing the Fourier transform of the velocity-velocity auto-correlation function sampled at 100~K. \Figref{fig:md17_vibrational_dos_300K_950_vs_128} and \figref{fig:md17_vibrational_dos_100K_950_vs_128} compare power spectra obtained by running simulations with forces from interatomic potentials fine-tuned on 128 and 950 energy and atomic force labels.

\end{document}